%% file: main.tex
\documentclass[aps, preprint, superscriptaddress,longbibliography]{revtex4-2}
\usepackage{graphicx} 
\usepackage{graphicx}
\usepackage{physics}
\usepackage{amsmath,amssymb,amsfonts,mathtools}
\usepackage{fancyhdr}
\usepackage{comment}
\usepackage[absolute,overlay]{textpos}

\newcommand\identity{1\kern-0.25em\text{l}}

\begin{document}
\title{Revealing spoofing of quantum illumination using entanglement}
\author{Jonathan N. Blakely}
\author{Shawn D. Pethel}
\author{Kenneth R. Stewart}
\address{U. S. Army DEVCOM Aviation \& Missile Center, Redstone Arsenal, Alabama 35898, USA}
\author{Kurt Jacobs}
\address{U. S. Army DEVCOM Army Research Laboratory, Adelphi, Maryland 20783, USA}
\begin{abstract}
    Several \lq quantum radar\rq\ concepts have been proposed that exploit the entanglement found in two-mode squeezed vacuum states of the electromagnetic field, the most prominent being radar based on quantum illumination. Classical radars are sometimes required to distinguish between true echos of their transmitted signals and signals generated by interferors or spoofers. How vulnerable to spoofing is quantum illumination? We analyze the scenario of a radar operator trying to detect the presence of a ‘classical’ spoofer employing a measure-and-prepare strategy against a quantum radar. We consider two spoofing strategies - (1) direct detection and number state preparation, and (2) heterodyne detection and coherent state preparation. In each case, the radar
operator performs a hypothesis test to decide if received pulses are true returns or spoofs. Since the spoofer can not reproduce the entanglement with modes retained by the radar operator, both approaches to spoofing are to some extent detectable. We quantify the effectiveness of the spoof in terms of the fidelity between the real return and the spoof return, and the probability of error in spoof detection. We find that in the absence of noise and loss, direct detection tends to produce spoofs with greater fidelity, which are therefore harder to detect. Moreover, this advantage survives the introduction of noise and loss into the model. Our results suggest that entanglement is a novel resource available to quantum radar for detecting spoofing. 
\end{abstract}

\maketitle

\section{Introduction}
Are there advantages to be gained in radar technology by employing quantum resources? Several \lq quantum radar\rq\, concepts have been proposed that exploit the entanglement found in two-mode squeezed vacuum states of the electromagnetic field \cite{luong2023quantum,karsa2023quantum, torrome2023advances,bischeltsrieder2024engineering}. Typically, signal modes are transmitted to probe a region of space, while idler modes are retained for use in determining if received radiation is a reflected pulse or radiation from some other source. The other source is usually environmental background noise, but that is not the only possibility.  Conventional radar technologies that operate in adversarial environments are often subject to spoofing, i.e., interfering signals generated by an adversary that are  designed to mislead the radar operator \cite{schleher1999electronic,genova2018electronic}. Spoofs are intended to fool the radar operator into wrongly estimating a target's location or speed, or to miss its presence altogether. It has recently been argued theoretically that a spoofer who employs a classical \lq measure-and-prepare\rq\, strategy to characterize an incoming radar pulse and generate a spoof is fundamentally limited by quantum noise \cite{blakely2022quantum}. Conversely, the imperfections of a spoof due to quantum noise can reveal the presence of the spoofer to the operator of a classical radar \cite{blakely2024revealing, wang2024quantum,espinoza2024quantum}. Does entanglement, the quantum resource exploited by quantum radar, present new options to radar technology subject to spoofing? In this article, that question is answered in the affirmative.

Among the various approaches to quantum radar, the one that has the most extensive theoretical support \cite{tan2008quantum,guha2009gaussian, nair2020fundamental, de2018minimum, bradshaw2021optimal,zhuang2022ultimate, karsa2023quantum, reichert2023quantum, angeletti2023microwave, wu2023entanglement, kronowetter2024imperfect,murchie2024object} and  experimental proof-of-concept demonstrations \cite{shapiro2020quantum, assouly2023quantum} is that based on a sensing protocol known as quantum illumination (QI). This protocol uses entanglement between modes (conventionally called signal and idler) generated by broadband spontaneous parametric down conversion to detect a target that is weakly reflecting and bathed in background thermal radiation, and measure its range, angle, and velocity  better than any classical radar can \cite{zhuang2022ultimate, wu2023entanglement, reichert2022quantum}. Interestingly, the quantum advantage of this protocol is greatest when loss and noise are sufficient to destroy the entanglement, leaving only its remnant, a weak classical correlation, at the receiver.

This article examines the basic physics of spoofing of a radar based on the entangled state of the electromagnetic field used in QI and other forms of quantum radar. Importantly, we show that entanglement provides a physical basis for spoof detection, similar to the role of quantum noise in previous studies \cite{blakely2022quantum, blakely2024revealing, wang2024quantum,espinoza2024quantum}.  Specifically, we analyze the scenario of a \lq classical\rq\, spoofer employing a measure-and-prepare strategy against a quantum radar. The radar operator performs a hypothesis test to decide if received pulses are true returns or spoofs. The null hypothesis, $H_0$, holds that the transmitted pulse is reflected by the target and returns to the radar operator. The alternative hypothesis, $H_1$, holds that spoofer receives the pulse, performs a measurement to estimate its quantum state, and transmits a corresponding spoof, which is received by the radar. It is assumed that the pulse received by the radar contains no echo of the transmitted signal modes (or skin return) \cite{blakely2022quantum}. Fundamentally, the entanglement between the signal modes and the idler modes cannot possibly be reproduced by the classical spoofer and, thus, provides a physical basis for discriminating between hypotheses. 

The QI protocol (and other quantum radar schemes) makes use of the entangled signal and idler fields produced by broadband, continuous-wave (cw) pumped, spontaneous parametric down conversion \cite{shapiro2020quantum, karsa2023quantum}. The signal field pulse is represented by the positive frequency, Heisenberg-picture operator, $\hat{E}_Se^{-i\omega_S t}$, where $\omega_S$ is the center frequency and $\hat{E}_S$ is the $\sqrt{\text{photons/s}}$-units, baseband, field operator. Similarly, the idler field is represented by the operator  $\hat{E}_Ie^{-i\omega_I t}$, with center frequency $\omega_I$ and baseband operator $\hat{E}_I$. The signal and idler fields are chopped into pulses of duration $T$ so their respective baseband operators can be conveniently expressed as Fourier series. Thus, 
\begin{align}
\hat{E}_S = \sum_{m=-\infty}^\infty \hat{a}_{S_{m}} \frac{e^{-i2\pi m t/T}}{\sqrt{T}},
\label{signal_field_op}
\end{align}
and
\begin{align}
\hat{E}_I = \sum_{m=-\infty}^\infty \hat{a}_{I_{m}} \frac{e^{-i2\pi m t/T}}{\sqrt{T}}.
\label{idler_field_op}
\end{align}
Here $\hat{a}_{S_{m}}$ is the annihilation operator for a single mode of the signal field with frequency $\omega_S+2\pi m/T$, and $\hat{a}_{I_{m}}$ is the annihilation operator for a single mode of the idler field with frequency $\omega_I+2\pi m/T$. It is assumed that only modes within bandwidth $W$ of $\omega_S$ and $\omega_I$ are excited. Thus, the pulses have a time-bandwidth product $M=TW$. Within the active bandwidth $W$ (i.e., where $m= -(M-1)/2,...,(M-1)/2)$, the $m^\text{th}$ signal mode and the $m^\text{th}$ idler mode are in the two-mode, squeezed vacuum state
\begin{align}
    \ket{\psi}_m=\sum_{n=0}^\infty \sqrt{\frac{N^n}{(1+N)^{1+n}}} \ket{n}_{I_m} \ket{n}_{S_m}.
    \label{TMSV_Fock}
\end{align}
In this state, each individual mode has mean photon number $N$. For simplicity, we assume this is true for any mode within the active bandwidth, i.e., the signal and idler fields have top-hat-shaped fluorescence spectra. The modes are correlated such that $\langle \hat{a}_{S_m}\hat{a}_{I_m}\rangle = \sqrt{N(N+1)}$ and $\langle\hat{N}_{I_m} \hat{N}_{S_m}\rangle =N+2N^2$ for any $m$, where $\hat{N}_{S_m}$ and $\hat{N}_{I_m}$ are the number operators for the $m^\text{th}$ signal and idler mode, respectively. The density operator for the total field is
\begin{align}
    \hat{\rho} = \otimes_{m=1}^M \ket{\psi}_m \prescript{}{m}{\bra{\psi}} ,
    \label{intial_QI_signal}
\end{align}
where the unexcited modes are suppressed. The signal beam (i.e., the field consisting of the sum of all signal modes) is used to interrogate a region of space, either to detect a target or to determine a previously detected target's range, velocity, angle, etc. The idler beam is retained for use in discriminating between a true echo off a target and radiation from the thermal background (or a spoof pulse, as described below). Broadband, two-mode, squeezed vacuum states are optimal for QI target detection \cite{nair2020fundamental, de2018minimum, bradshaw2021optimal}.

A classical spoofer intercepts a signal pulse, makes a measurement to inform state estimation, prepares a corresponding spoof pulse, and transmits it to the radar receiver. Different types of measurement made by the spoofer provide more or less information for preparing spoofs. 
To specify fundamental limits on the spoofer 
 we would assume the spoofer employs the quantum optimal strategy for estimating the state of the pulse from a single measurement \cite{blakely2022quantum}. However, we do not know an optimal measure-and-prepare strategy for this scenario. Therefore, we consider two specific spoofing strategies based on common measurements employed in quantum optics, namely, direct detection and heterodyne detection \cite{ou2017quantum}. In the first strategy, we assume the spoofer performs mode-wise direct detection and generates a spoof pulse  in a multi-mode number state. In the second strategy, we assume the spoofer performs mode-wise, heterodyne detection and generates a spoof pulse in a multi-mode coherent state \cite{jacobs2014quantum}. In analyzing these spoofing strategies, we first consider a toy model of spoofing that neglects effects of noise and loss. The value of this model is that it makes clear the purely quantum resource (entanglement) employed for spoof detection. In Secs. \ref{direct_spoof} and \ref{hetero_spoof}, the two potential spoofing strategies are analyzed within this idealized framework. We find that the former method produces spoof pulses that are harder to distinguish from true reflections than does the latter. In Sec. \ref{noise_channel}, we add noise and loss to the model of both strategies. We find that the advantage of the former method of spoofing over the latter for producing credible spoofs can survive the introduction of these effects. Altogether, our results show (1) that entanglement can be used to detect spoofing in QI, and (2) that mode-wise direct detection and multi-mode number state preparation is the best known strategy for a spoofer to avoid such detection.

\section{Spoofing with Direct Detection and Number State Preparation}
\label{direct_spoof}
The first spoofing strategy we analyze employs mode-wise direct detection followed by preparation of a multi-mode number state. We assume detection is performed using a microwave, square-law detector that conforms approximately to conventional quantum photo-detection theory \cite{haus2012electromagnetic}. This spoofing strategy is of interest for two reasons. First, it exploits the perfect correlation in photon number between the two modes of each two-mode, squeezed vacuum state. Secondly, mode-wise direct-detection is conceptually similar to the functioning of a conventional spectrum analyzer insofar as it measures the magnitude of the signal at each frequency. In general, number states with more than one photon are not easy to prepare. However, in QI the mean photon number per mode is small compared to one. So in practice, the number states being generated would likely have either zero or one photon. 

Spoof detection involves making a decision, informed by collected data, whether a received pulse is a true return or a spoof. As stated above, we frame this decision as a hypothesis test \cite{1976quantum}. The null hypothesis, $H_0$, holds that the radar receives a true reflection of the transmitted signal-mode pulse, i.e., the received field is in the same state as the signal field apart from effects of noise and loss, which are neglected until Sec. \ref{noise_channel}. Introducing a new baseband operator to represent the received field
\begin{align}
\hat{E}_R = \sum_{m=-\infty}^\infty \hat{a}_{R_{m}} \frac{e^{-i2\pi m t/T}}{\sqrt{T}}.
\end{align}
The noise-free, loss-free 
 or \lq identity\rq\, channel  from transmission of the signal field to arrival of the received field is described by the trivial relationship 
\begin{align}
    \hat{a}_{R_m}=\hat{a}_{S_m},
\end{align}
for all $m$. (In Sec. \ref{noise_channel}, we will replace this identity channel with a more realistic noise-loss channel.) The density operator representing the two fields at the radar receiver (i.e., retained idler $\hat{E}_I$ and received radiation $\hat{E}_R$) is simply
\begin{align}
    \hat{\rho}_{H_0} = \otimes_{m=1}^M \ket{\psi}_m \prescript{}{m}{\bra{\psi}} ,
    \label{direct_null}
\end{align}
where $\ket{\psi}$ is the two-mode, squeezed vacuum state defined by Eq. \ref{TMSV_Fock}. 

Under the alternative hypothesis, $H_1$, the spoofer intercepts the pulse, makes a mode-wise measurement of photon number, and prepares a spoof pulse in which each mode is in a number state with the corresponding measured number of photons. For clarity, we describe the spoofer's actions in terms of their effects on the $m^\text{th}$ received mode. The spoofer has no access to the corresponding idler mode. Thus, the spoofer's single-mode measurement statistics are determined by the reduced density operator $\trace_{I_m} (\ket{\psi}_m \prescript{}{m}{\bra{\psi}})$. Specifically, direct detection produces measurement outcome $n_m$ with probablity
\begin{align}
    &P(n_m) =\trace \Bigl [\ket{n_m}_S\prescript{}{S}{\bra{n_m}}\trace_{I_m} (\ket{\psi}_m \prescript{}{m}{\bra{\psi}}) \Bigr] =\frac{1}{1+N} \left(\frac{N}{1+N}\right)^{n_m}  
    \label{Plank_distribution}.
\end{align}
Supposing the spoofer has obtained outcome $n_m$, the $m^\text{th}$ mode of the spoof field is prepared in the number state $\ket{n_m}$, which becomes the $m^\text{th}$ mode received by the radar. 

We assume the radar operator knows the spoofer would make a photon number measurement of the $m^\text{th}$ signal mode, but does not know the outcome of the measurement. The photon numbers of the signal and idler modes of a two-mode, squeezed vacuum are perfectly correlated. Measurements of their individual photon numbers produce identical outcomes. The state of the $m^\text{th}$ retained idler mode conditioned on $n_m$, the outcome of the spoofer's measurement, is then
\begin{align}
    \hat{\rho}_{I_m|n_m}&= \ket{n_m}_{I_m}\prescript{}{I_m}{\bra{n_m}}.
\end{align}
Since the radar operator does not know the spoofer's measurement outcome, the measurement statistics at the radar receiver are determined by the two-mode state that is a statistical mixture of all possible outcomes. The density operator for the $m^\text{th}$ idler-received mode pair is
\begin{align}
    \hat{\rho}_{I_m,R_m|H_1}&=\sum_{n_m=0}^\infty  P(n_m) \ket{n_m}_{R_m}\prescript{}{{R_m}}{\bra{n_m}}\otimes \hat{\rho}_{I_m|n_m}\label{direct_alt}\\
    &=\sum_{n_m=0}^\infty \frac{1}{1+N} \left(\frac{N}{1+N}\right)^{n_m} \;\;\;\ket{n_m}_{I_m}\prescript{}{I_m}{\bra{n_m}} \otimes\ket{n_m}_{R_m}\prescript{}{{R_m}}{\bra{n_m}}.
\end{align}
This is a mixture (as opposed to a  superposition) of product states (density operators), so it is not entangled \cite{wilde2013quantum}. But there is a classical correlation of photon number between the two modes such that $\langle\hat{N}_{I_m} \hat{N}_{S_m}\rangle =N+2N^2$, just like the two-mode, squeezed vacuum. If the measurement outcome for the $m^\text{th}$ mode is $n_m$, then the multi-mode density operator for the alternate hypothesis, $H_1$, is 
\begin{align}
    &\hat{\rho}_{H_1} = \otimes_{m=1}^M \;\hat{\rho}_{I_m,R_m|H_1} 
        \label{direct_alt}
\end{align}

\begin{figure}[tbh]
\includegraphics{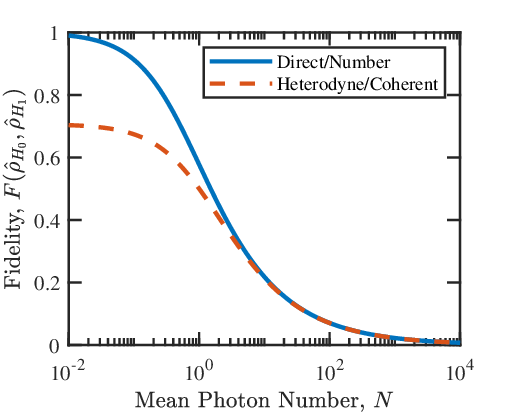}
\caption{Fidelity of the density operators for the null and alternate hypotheses as a function of mean photon number for a single mode (i.e., $M=1$) and a spoofer who employs direct detection and number states (solid blue line) and heterodyne detection and coherent states (dashed red line). }
\label{fig:fidelity_comparison}
\end{figure}

Any difference between $\hat{\rho}_{H_0}$ and $\hat{\rho}_{H_1}$ constitutes a physical basis for discrimination between the hypotheses. Such a difference can be quantified in terms of quantum fidelity. The fidelity of these density operators is
\begin{align}
F(\hat{\rho}_{H_0}, \hat{\rho}_{H_1}) &=\sqrt{\prescript{}{1}{\bra{\psi}}\prescript{}{2}{\bra{\psi}}...\prescript{}{M}{\bra{\psi}} \hat{\rho}_{H_1} \ket{\psi}_1\ket{\psi}_2...\ket{\psi}_M}, \\
&= \left ( \frac{1}{\sqrt{2N+1}} \right )^M.
\label{Direct_fidelity}
\end{align}
The fidelity $F$ for a single mode (or $M=1$) is shown in Fig. \ref{fig:fidelity_comparison} (solid blue line) as a function of the single mode, mean photon number $N$. Notably, as $N$ tends towards zero, the fidelity tends towards unity. In this limit, both density operators, Eqs. \ref{direct_null} and \ref{direct_alt}, approach vacuum states and spoofing becomes undetectable. In the opposite limit where $N$ tends towards infinity and the null hypothesis density operator, Eq. \ref{direct_null}, describes a maximally entangled state, the fidelity tends towards zero. As a result, the stronger the transmitted power, the greater physical basis for discrimination.

\begin{figure}[tbh]
\includegraphics{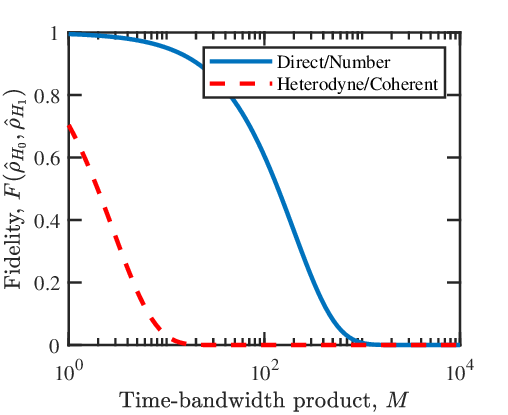}
\caption{Fidelity of the density operators for the null and alternate hypotheses as a function of time-bandwidth product $M$ for $N=0.01$ and a spoofer who employs direct detection and number states (solid blue line) and heterodyne detection and coherent states (dashed red line). }
\label{fig:tbw_comparison}
\end{figure}

The fidelity is shown again in Fig. \ref{fig:tbw_comparison} (solid blue line) but as a function of the time-bandwidth product $M$ when $N=0.01$, a typical value in the QI literature \cite{angeletti2023microwave, shi2024optimal, shapiro2020quantum, karsa2020quantum}. For such a weak signal, the fidelity is near unity when $M$ is less than about $10^2$, i.e., where $MN \approx 1$, but it drops off rapidly for larger values of $M$. From Eq. \ref{Direct_fidelity}, it can be seen that, for any finite $N$, the fidelity approaches zero as $M \to \infty$. Thus, the larger the time-bandwidth product, the greater physical basis for discrimination.

\begin{figure}[tbh]
\includegraphics{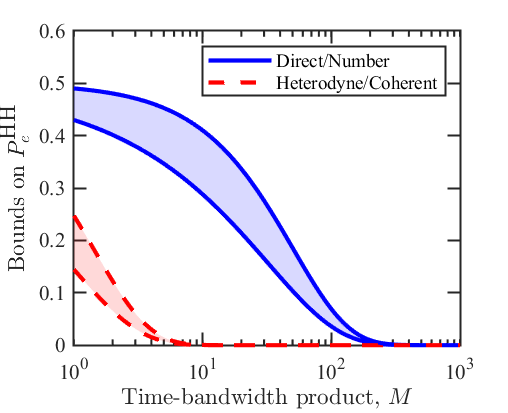}
\caption{Upper and lower bounds on the optimal error probability in discriminating a spoofer who employs direct detection and number states (solid blue line) and a spoofer who employs heterodyne detection and coherent states (dashed red line). These results assume equal prior probabilities and a single-mode, mean photon number $N=0.01$. }
\label{fig:popt_comparison}
\end{figure}

Our conclusions based on fidelity can be confirmed by consideration of the probability $P_e$ that the radar operator errs in choosing the true hypothesis. This probability includes both errors of selecting the null hypothesis when the alternate hypothesis is true, and errors of selecting the alternate hypothesis when the null hypothesis is true. Assuming equal Bayesian prior probabilities for the two hypotheses, and equal costs for both types of error, quantum mechanics requires the error probability to be bounded by the inequality
\begin{align}
P_e \le P_e^\text{HH} \equiv \frac{1}{2} \left(1- \frac{1}{2}||\hat{\rho}_{H_1}-\hat{\rho}_{H_0}||_1\right),
\label{P_err}
\end{align}
 and where $|| \cdot ||_1$ denotes the trace norm \cite{helstrom1967detection, 1976quantum}. Equality is achieved by a receiver that realizes the Helstrom-Holevo test (hence the superscript HH in Eq. \ref{P_err}), a test that involves a projective measurement onto the positive part of $\hat{\rho}_{H_1}-\hat{\rho}_{H_0}$. The optimal error probability $P_e^\text{HH}$ is generally difficult to calculate, but given Eq. \ref{Direct_fidelity} and the fact that $\hat{\rho}_{H_0}$ is a pure state, the $P_e^\text{HH}$ is bounded by the inequalities \cite{nielsen2010quantum}.
\begin{align}
    \frac{1}{2}\left( 1-\sqrt{1-\left(\frac{1}{1+N+N^2} \right )^M}\right)\le P_e^\text{HH} \le \frac{1}{2} \left( \frac{1}{1+N+N^2} \right )^M .
\end{align}
 These bounds are shown in Fig. \ref{fig:popt_comparison} as functions of $M$. In the limiting case of $N=0$, the received modes are in vacuum states under both hypotheses. Then it is impossible to discriminate, so $P_e^\text{HH} =1/2$, which is just the assumed prior probability of spoofing. Thus, the combination of mode-wise direct detection and multi-mode number state generation produces asymptotically perfect spoof pulses. On the other hand, for $N \to \infty$, $P_e^\text{HH} \to 0$. Similarly, for any finite $N$ and $M \to \infty$, $P_e^\text{HH} \to 0$. The brighter the pulses and the larger the time-bandwidth product, the easier it is to distinguish the presence or absence of entanglement between the idler and the received light. 

\begin{figure}[tbh]
\includegraphics{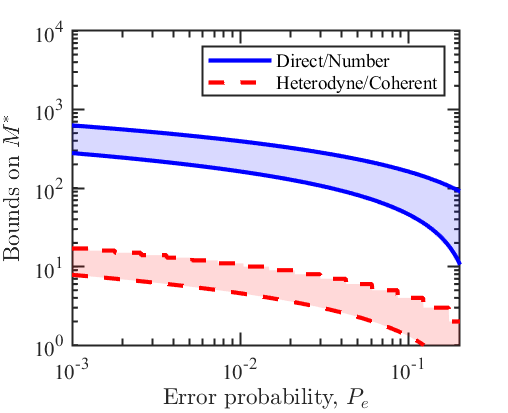}
\caption{Upper and lower bounds on time-bandwidth product required to achieve probability of error $P_e$ using optimal detection with $N=0.01$.}
\label{fig:stat_comp}
\end{figure}

 In quantum hypothesis testing generally, the fidelity $F(\hat{\rho}_{H_0}, \hat{\rho}_{H_1})$ can be used to bound the minimum number of samples required to obtain a desired probability of error $P_e$ \cite{cheng2024sample}. Here samples are identical copies of the density operator being tested. In the current problem, since each mode is independent and identical, the number of modes $M$, which is also the time-bandwidth product, is formally equivalent to a number of repeated samples of a single mode. Thus, if the minimum time-bandwidth product required to obtain a desired error probability is denoted $M^*(P_e)$, then given Eq. \ref{Direct_fidelity}, it follows that 
 \begin{align}
     -\frac{\ln\left(4P_e\right)}{\ln \left ( 1+N+N^2 \right )} \le M^*(P_e) \le \left\lceil -\frac{\ln\left(2P_e\right)}{\frac{1}{2}\ln \left ( 1+N+N^2 \right )} \right\rceil,
\end{align}
where again equal prior probabilities are assumed \cite{cheng2024sample}. For example, the bounds are shown as blue lines in Fig. \ref{fig:stat_comp} with $N=0.01$. Interestingly, the upper bound can be attained by a radar that implements mode-wise, Fuchs-Caves measurements \cite{fuchs1995mathematical,cheng2024sample}. Even better performance may require a collective measurement on all $M$ modes. In any case, these bounds are consistent with the earlier observation that increasing the brightness of a pulse reduces the required time-bandwidth product to achieve a desired probability of error, particularly in the regime where $N<<1$. 

It is worth pausing to consider how the results presented so far would be different if, instead of the entangled mode pairs of the two-mode, squeezed vacuum, we had begun with classically correlated mode pairs. Such a state could be represented by the density operator of the form 
\begin{align}
    &\hat{\rho} =\otimes_{m=1}^M \sum_{n_m=0}^\infty P(n_m) \;\ket{n_m}_{I_m}\prescript{}{I_m}{\bra{n_m}} \otimes\ket{n_m}_{S_m}\prescript{}{{S_m}}{\bra{n_m}} 
    ,
\end{align}
where $P(n_m)$ is a probability distribution over number states. Consider the fidelity of $\hat{\rho}_{H_0}$ and $\hat{\rho}_{H_1}$. It is straightforward to show that the same reasoning as above leads in this case to $F(\hat{\rho}_{H_0}, \hat{\rho}_{H_1})=1$. Thus, in this simple model with no loss or noise, the spoofer can perfectly reproduce a signal exhibiting only classical correlations. In contrast, entanglement is a quantum resource that can reveal spoofing.

\begin{figure}[tbh]
\includegraphics{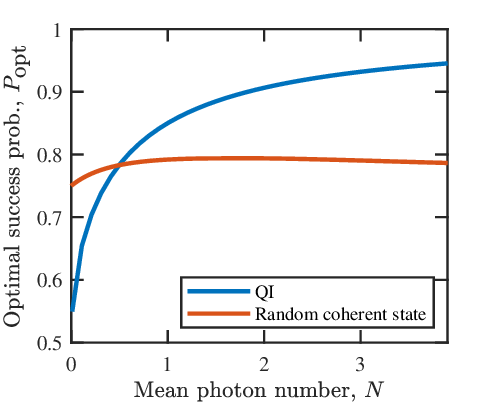}
\caption{Optimal probability of success in choosing the true hypothesis as a function of mean photon number for QI (blue line) and random coherent states (red line). }
\label{fig:TMSV_coh}
\end{figure}

A final interesting comparison can be made between spoof detection using the QI signal and spoof detection using coherent states with random amplitudes. In Ref. \cite{blakely2022quantum}, the optimal probability of successful discrimination $P_\text{opt} = 1- P_e^\text{HH}$ was numerically estimated when the transmitted signal is a single mode coherent state whose amplitude was drawn from a Gaussian distribution. The estimate was arrived at by (1) truncating the number state representations of the density operators $\hat{\rho}_{H_0}$ and $\hat{\rho}_{H_1}$ at a large, but finite photon number, (2) solving numerically for the eigenvalues of $\hat{\rho}_{H_1}-\hat{\rho}_{H_0}$, and (3) summing the absolute values of the eigenvalues to approximate the trace norm in Eq. \ref{P_err}. The result of the calculation is reproduced here as the red line in Fig. \ref{fig:TMSV_coh}.  By the same method, we estimate $P_\text{opt}$ as a function of the mean photon number $N$ for a single-mode QI pulse subject to spoofing using direct detection and number state preparation. The result appears as the blue line in Fig. \ref{fig:TMSV_coh}, which exceeds the red line for mean photon numbers greater than about 0.5. Within the limitations of this simple model, it can be concluded that the entanglement in the QI signal provides a better basis for spoof detection than does the quantum noise of randomly drawn coherent states.

Another difference between these approaches to spoof detection is important to note. When a state is randomly drawn from some set of possible signals, the parameters that define that state function as a secret key. The spoofer's task is to guess the secret key. Knowledge of the key (e.g., the complex amplitude of a randomly drawn coherent state) enables perfect spoofing. In contrast, the idler modes function as the key in the QI signal. But in this case, knowledge of the parameters that characterize the signal (i.e., $N$, $M$, $\omega_S$, $\omega_I$, and $T$) do not enable perfect spoofing. It seems that in order to create a perfect spoof, the spoofer would either need to gain physical possession of the idler modes themselves or resort to non-classical operations to transfer entanglement from the signal modes to the spoof modes. Thus, entanglement provides a qualitatively different physical basis for revealing spoofing.

\section{Spoofing with Heterodyne Detection and Coherent State Generation}
\label{hetero_spoof}
The second spoofing strategy we analyze employs mode-wise, heterodyne detection followed by preparation of a multi-mode coherent state. This strategy is of interest because it is likely to be more easily implemented in practice than the previous strategy. Firstly, coherent states are much easier to generate than are number states. Secondly, single-mode, heterodyne detection is functionally equivalent to an existing, conventional receiver architecture, namely, IQ demodulation. A single-mode, heterodyne measurement outcome consists of two real numbers or, equivalently, one complex amplitude. Mode-wise, heterodyne detection resembles the function of a vector signal analyzer, which measures both the amplitude and phase of an input signal at each frequency. 

In the spoof detection hypothesis test for this strategy, the null hypothesis, $H_0$, does not involve the spoofer at all. Therefore, Eq. \ref{direct_null} again correctly represents the idler and received fields at the radar receiver. As in Sec. \ref{direct_spoof}, under the alternative hypothesis, $H_1$, the spoofer receives the signal mode, but has no access to the idler mode. Thus, the spoofer's single-mode measurement statistics for the $m^\text{th}$ mode are determined by the reduced density operator $\trace_{I_m} (\ket{\psi}_m \prescript{}{m}{\bra{\psi}})$. The outcome of a single-mode, heterodyne measurement is a pair of real numbers that are the real and imaginary parts of a complex amplitude. An idealized heterodyne measurement is represented by a positive, operator-valued measure with effect operators
\begin{align}
    \hat{M}_{\alpha} \equiv \frac{1}{\sqrt{2\pi}} \ket{\alpha}_S \prescript{}{S}{\bra{\alpha}},
    \label{het_POVM}
\end{align}
where $\ket{\alpha}_S$ is a coherent state of the signal mode with mean amplitude $\alpha \in \mathbb{C}$.
The probability density associated with a heterodyne measurement of the $m^\text{th}$ signal mode with outcome $\alpha_m$ is
\begin{align}
    p(\alpha_m) =\trace{\left(\hat{M}_{\alpha_m}\trace_I (\ket{\psi}\bra{\psi})\hat{M}_{\alpha_m}^\dagger\right)}=\frac{e^{\frac{-|\alpha_m|^2}{N+1}}}{2\pi(N+1)}.
\end{align}
Supposing the spoofer has obtained outcome $\alpha_m$, the $m^\text{th}$ mode of the spoof field is prepared in the coherent state $\ket{\alpha_m}$, which becomes the $m^\text{th}$ mode received by the radar.

The radar operator knows the spoofer would make a heterodyne measurement of the signal, but does not know $\alpha_m$, the outcome of the measurement. The state of the $m^\text{th}$ mode of the retained idler conditioned on $\alpha_m$ is
\begin{align}
    \hat{\rho}_{I_m|\alpha_m}=\trace_{S_m} \left( \frac{\hat{M}_{\alpha_m}\ket{\psi}\bra{\psi}\hat{M}_{\alpha_m}^\dagger}{p(\alpha_m)}  \right) =\ket{\sqrt{\frac{N}{1+N}}\alpha_m^*}_{I_m}\prescript{}{I_m}{\bra{\sqrt{\frac{N}{1+N}}\alpha_m^*}}.
\end{align}
Since the radar operator does not know the measurement outcome, measurement statistics at the radar receiver are determined by the two-mode state that is a statistical mixture of all possible outcomes. The density operator for the $m^\text{th}$ idler-received mode pair is
\begin{align}
    \hat{\rho}_{I_m,R_m|H_1} &=\int d^2 \alpha_m \; p(\alpha_i) \ket{\alpha_m}_{R_m}\prescript{}{{R_m}}{\bra{\alpha_m}}\otimes \hat{\rho}_{I_m|\alpha_m}\\
    &=\int d^2 \alpha_m \; \frac{1}{2\pi(N+1)}e^{\frac{-|\alpha_m|^2}{N+1}} \ket{\alpha_m}_{R_m}\prescript{}{{R_m}}{\bra{\alpha_m}}\otimes \ket{\sqrt{\frac{N}{1+N}}\alpha_i^*}_{I_m}\prescript{}{I_m}{\bra{\sqrt{\frac{N}{1+N}}\alpha_m^*}}.
\end{align}
This is a mixture of product states, so it is not entangled, but classically correlated. If the measurement outcome for the $m^\text{th}$ mode is $\alpha_m$, then the multi-mode density operator for the alternate hypothesis, $H_1$, is
\begin{align}
    &\hat{\rho}_{H_1} = \otimes_{m=1}^M \;\hat{\rho}_{I_m,R_m|H_1} 
        \label{hetero_alt}
\end{align}

The fidelity of $\hat{\rho}_{H_0}$ and $\hat{\rho}_{H_1}$ is
\begin{align}
F(\hat{\rho}_{H_0}, \hat{\rho}_{H_1}) &=\sqrt{\prescript{}{1}{\bra{\psi}}\prescript{}{2}{\bra{\psi}}...\prescript{}{M}{\bra{\psi}} \hat{\rho}_{H_1} \ket{\psi}_1\ket{\psi}_2...\ket{\psi}_M} \\
&= \left ( \frac{1}{\sqrt{2N+2}} \right )^M.
\label{Hetero_fidelity}
\end{align}
Figure \ref{fig:fidelity_comparison} compares the fidelity $F$ in the single mode case (or $M=1$) as a function of the mean photon number $N$ for a spoofer who employs direct detection and number states (solid blue line) to that of a spoofer who uses heterodyne detection and coherent states (dashed red line). For the latter, the limit of $N\to 0$ gives $F\to 1/\sqrt{2}$ rather than zero. This reflects the weaker correlations in vacuum fluctuations that are present in measurements of complex amplitude than in photon number. Thus, this approach to spoofing cannot produce perfect spoofs at any power level. As $N$ gets large, the fidelity for heterodyne detection and coherent states approaches zero indicating, again, that the ability to detect spoofing increases with signal brightness. 

Figure \ref{fig:tbw_comparison} makes a similar comparison but with fidelity as a function of the time-bandwidth product $M$, and with $N=0.01$. Strikingly, the fidelity declines much more rapidly with increasing $M$ if the spoofer employs heterodyne detection and coherent states.

The bounds on the optimal error probability $P_e^\text{HH}$, assuming equal prior probabilities, introduced in Sec. \ref{direct_spoof}, are in this case 
\begin{align}
      \frac{1}{2}\left( 1-\sqrt{1-\left ( \frac{1}{2N+2} \right )^M}\right)\le P_e^\text{HH} \le \frac{1}{2} \left ( \frac{1}{2N+2} \right )^M .
\end{align}
In the limit where $N\to 0$, the lower bound is greater than the prior probability of 1/2 for any finite $M$. Thus, the combination of mode-wise heterodyne detection and multi-mode coherent state generation does not produce asymptotically perfect spoof pulses. Figure \ref{fig:popt_comparison} compares the bounds for heterodyne detection and coherent states to the bounds for direct detection and number states as functions of the time-bandwidth product and $N=0.01$. Since the lower bound on the former is above the upper bound on the latter for all $M$, the figure shows clearly how much easier spoofs based on heterodyne detection and coherent states are to detect than those based on direct detection and number states. A similar observation can be drawn from the bounds on the minimum time-bandwidth product to achieve a desired error probability, which in this case are
\begin{align}
     -\frac{\ln\left(4P_e\right)}{\ln \left ( 2N+2 \right )} \le M^*(P_e) \le \left\lceil -\frac{\ln\left(2P_e\right)}{\frac{1}{2}\ln \left (2N+2 \right )} \right\rceil.
\end{align}
In conclusion, the ability to detect a spoofer who used heterodyne detection and coherent states increases with signal brightness and bandwidth. But at any brightness and bandwidth, it is always more difficult to detect a spoofer who uses direct detection and number states.

\section{Noise-loss Channel Model}
\label{noise_channel}

The simple model of a spoofing scenario introduced in the previous section led to the  conclusion that entanglement in QI may be a useful resource for revealing spoofing. It was also shown that spoofs based on direct detection and number state generation were more difficult to detect than spoofs based on heterodyne detection and coherent state generation. In this section, we bring a greater degree of realism into the analysis by adding noise and loss to the spoofing model. To introduce noise and loss, we suppose the transmitted signal modes pass through noise-loss channels in propagation to the target and back, as detailed previously \cite{blakely2024revealing}. 

A noise-loss channel, denoted $\mathcal{L}_{\tau, N_i}$, is characterized by two parameters, the transmissivity $\tau$ and the mean photon number $N_i$ of the associated noise. Such a channel can be viewed as a beam splitter with thermal noise of mean photon number $N_i$ entering one input port to be combined with the channel input at the other port. The mean photon number at the channel output due purely to noise  is then $N_o \equiv N_i (1-\tau)$. In what follows, we set $N_i = N_o/(1-\tau)$ so that the mean photon number at the output due to noise is $N_o$ independent of $\tau$.

 Under the null hypothesis, a single channel $\mathcal{L}_{\tau, N_i}^S$ represents the entire round trip from transmitter to target to receiver. This channel lumps together all noise sources and loss mechanisms in effect during transmission, propagation, reflection, and reception into a single Gaussian noise source with mean photon number $N_i$ and an overall transmissivity $\tau$. The $S$ superscript here indicates that only the signal mode of each mode pair passes through the channel. For simplicity, we assume all modes are subject to the same noise and loss. The idler modes are assumed to be stored without any loss or added noise. This process transforms the initial idler and signal fields in state $\hat{\rho}$ as defined in Eq. \ref{intial_QI_signal} into idler and received fields in the state
\begin{align}
    &\hat{\rho}_{H_0}=\otimes_{i=1}^M \mathcal{L}_{\tau, N_i}^{S_i} \left(\ket{\psi}_i \prescript{}{i}{\bra{\psi}}\right)
     \label{DD_rho_0}.
\end{align}
A number state representation of the effect of the channel, used in numerical calculations described below, is given in the appendix.

Under the alternate hypothesis, the round trip is modelled by three operations. First, the transmitted signal modes pass through a noise-loss channel $\mathcal{L}_{\sqrt{\tau}, N_i}^S$. Note the transmissivity in this case is $\sqrt{\tau}$, representing half the loss of a full round trip. Second, the spoofer performs a measurement and generates a new state accordingly. Third, the new state passes through the channel $\mathcal{L}_{\sqrt{\tau}, N_i}^S$ a second time while the pulse travels back to the radar receiver. If the spoofer simply reflected the transmitted state, the result would be indistinguishable from the null hypothesis, i.e., $\mathcal{L}_{\sqrt{\tau}, N_i}^S\left(\mathcal{L}_{\sqrt{\tau}, N_i}^S\left(\hat{\rho}\right)\right)=\mathcal{L}_{\tau, N_i}^S\left(\hat{\rho}\right)$. We next examine the effects of the spoofer's actions under the two spoofing strategies introduced above.

\subsection{Direct Detection and Number State Generation}
Employing the strategy of Sec. \ref{direct_spoof}, the spoofer receives the state $\otimes_{m=1}^M\mathcal{L}_{\sqrt{\tau}, N_i}^{S_m}\left(\hat{\rho}\right)$ and performs mode-wise direct detection and prepares a multi-mode number state in accord with the measurement outcomes. 
Ideal direct detection of the $m^\text{th}$ signal mode produces an outcome $n_m$ with probability
\begin{align}
    P(n_m)     &= \trace \left[\left(\identity_{I_m} \otimes\ket{n_m}_{S_m}\prescript{}{S_m}{\bra{n_m}}\right)\mathcal{L}_{\sqrt{\tau}, N_i}^{S_m}\left(\hat{\rho}\right) \right]
\end{align}The post-measurement state of the $m^\text{th}$ idler-signal mode pair conditioned on the measurement outcome $n_m$ is
\begin{align}
    \hat{\rho}_{n_m}&=\frac{\ket{n_m}_{S_m}\prescript{}{S_m}{\bra{n_m}}\mathcal{L}_{\sqrt{\tau}, N_i}^{S_m}\left(\hat{\rho}\right)\ket{n_m}_{S_m}\prescript{}{S_m}{\bra{n_m}}}{P(n_m)} 
\end{align}
 This expression represents a signal mode projected into a number state by a measurement and an idler state projected into an imperfectly correlated mixture of number states. The spoofer transmits a field with each mode in a number state such that the $m^\text{th}$ mode has photon number $n_m$. The radar operator does not know the spoofer's direct detection outcome, only the measurement statistics. Then the joint density operator for the idler modes and the modes that will eventually be received for the field as it leaves the spoofer is
\begin{align}
    \otimes_{m=1}^M\sum_{n_m=0}^\infty  P(n_m)\hat{\rho}_{n_m} 
\end{align}
The field emitted by the spoofer passes through the channel $\mathcal{L}_{\sqrt{\tau}, N_i}^S$ on its way to the radar receiver where the final density operator is
\begin{align}
    \hat{\rho}_{H_1} &=  \otimes_{m=1}^M\mathcal{L}_{\sqrt{\tau}, N_i}^{S_m} \left( \sum_{n_m=0}^\infty  P(n_m)\hat{\rho}_{n_m}  \right) 
\label{DD_rho_1}
\end{align}

An exact expression for the fidelity  of $\hat{\rho}_{H_0}$ and $\hat{\rho}_{H_1}$ is not easily derived. So we resort to a numerical calculation at specific parameter sets. The estimate is arrived at by (1) truncating the number state representations of the density operators $\hat{\rho}_{H_0}$ and $\hat{\rho}_{H_1}$ at a large, but finite photon number, and (2) calculating the fidelity from the definition $F(\hat{\rho}_{H_0}, \hat{\rho}_{H_1}) \equiv \trace \sqrt{\sqrt{\hat{\rho}_{H_0}}\hat{\rho}_{H_1}\sqrt{\hat{\rho}_{H_0}}}$. A number state representation of the effect of the channel truncated at some large finite photon number is given in the appendix. The fidelity is first calculated for a single mode and then raised to the $M^\text{th}$ power since fidelity is generally multiplicative over tensor products \cite{fuchs1999cryptographic}. 

\begin{figure}[tbh]
\includegraphics{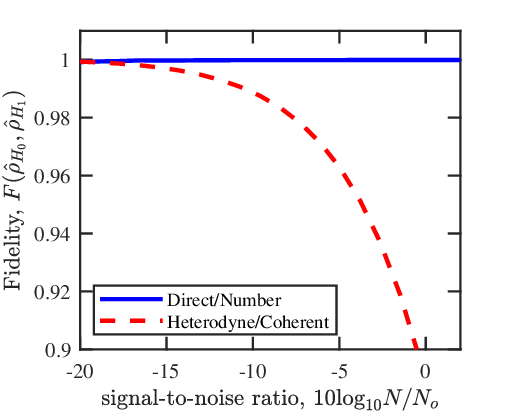}
\caption{The solid blue line indicates the fidelity for a spoofer using direct detection and number state preparation. The dashed red line indicates the fidelity for heterodyne detection and coherent state preparation. The mean photon number for both modes prior to the channel is $N=0.01$. The time-bandwidth product is $M=10^4$. The round trip transmissivity $\tau=10^{-6}$. The number state representations of the density operators used to calculate the fidelity were truncated at a maximum photon number of 34.}
\label{fig:fidelity_in_noise}
\end{figure}

Figure \ref{fig:fidelity_in_noise} shows results typical of all parameter values we have examined. In this case, $N = 0.01$ and $M=10^4$. We set $\tau = 10^{-6}$, as losses in radar are quite extreme. The number state representations of the density operators used to calculate the fidelity were truncated at a maximum photon number of 34. We consider $N_o$ values sampling a range of signal-to-noise ratios (defined as $N/N_o$) covering several orders of magnitude. The fidelity is very close to unity, indistinguishable within the resolution of the plot, over the entire range of signal-to-noise ratios. At this high level of loss, detection of a spoofer using this strategy is very difficult. A possible adjustment the radar operator could make to improve detection is to increase signal power. As Fig. \ref{fig:fidelity_comparison} indicates, increasing $N$ by an order of magnitude or more can dramatically reduce the fidelity of the spoof pulses. This may come at the cost of moving the radar out of the operating regime in which it has a quantum advantage in target detection. Perhaps there are scenarios where the operator temporarily increases signal power to test for spoofing before lowering the signal power for target detection. Also, Bayesian inference can be used to integrate information from many pulses to improve the ability to detect spoofs \cite{blakely2022quantum, blakely2024revealing}. In any case, we do not pursue these ideas further here. We now turn to the analysis of heterodyne detection and coherent states with noise and loss.

\subsection{Heterodyne Detection and Coherent State Generation}
Employing the strategy of Sec. \ref{hetero_spoof}, the spoofer receives the state $\otimes_{m=1}^M\mathcal{L}_{\sqrt{\tau}, N_i}^{S_m}\left(\hat{\rho}\right)$, performs mode-wise heterodyne detection, and prepares a multi-mode, coherent state in accord with the measurement outcomes. Ideal heterodyne detection of the $m^\text{th}$ signal mode produces an outcome $\alpha_m$ with probability density
\begin{align}
    p(\alpha_m)     &= \trace \left[\left(\identity_{I_m} \otimes\ket{\alpha_m}_{S_m}\prescript{}{S_m}{\bra{\alpha_m}}\right)\mathcal{L}_{\sqrt{\tau}, N_i}^{S_m}\left(\hat{\rho}\right) \right]
\end{align}The post-measurement state of the $m^\text{th}$ idler-signal mode pair conditioned on the measurement outcome $\alpha_m$ is
\begin{align}
    \hat{\rho}_{I_m|\alpha_m}&=\trace_{S_m} \left(\frac{\ket{\alpha_m}_{S_m}\prescript{}{S_m}{\bra{\alpha_m}}\mathcal{L}_{\sqrt{\tau}, N_i}^{S_m}\left(\hat{\rho}\right)\ket{\alpha_m}_{S_m}\prescript{}{S_m}{\bra{\alpha_m}}}{p(\alpha_m)} \right)
\end{align}
The spoofer transmits a field with each mode in a coherent state such that the $m^\text{th}$ mode has mean complex amplitude $\alpha_m$. The radar operator does not know the spoofer's heterodyne detection outcome, only the measurement statistics. Then the joint density operator for the idler modes and the modes that will eventually be received for the field as it leaves the spoofer is
\begin{align}
    \otimes_{m=1}^M\int d^2\alpha_m  p(\alpha_m)\hat{\rho}_{I_m|\alpha_m} \otimes \ket{\alpha_m}_{S_m}\prescript{}{S_m}{\bra{\alpha_m}}
\end{align}
The field emitted by the spoofer passes through the channel $\mathcal{L}_{\sqrt{\tau}, N_i}^{S_m}\left(\hat{\rho}\right)$ on its way to the radar receiver where the final density operator is
\begin{align}
    \hat{\rho}_{H_1} &=  \otimes_{m=1}^M \int d^2\alpha_m\;  p(\alpha_m)\hat{\rho}_{I_m|\alpha_m}\otimes \mathcal{L}_{\sqrt{\tau}, N_i}^{R_m} \left(\ket{\alpha_m}_{R_m}\prescript{}{R_m}{\bra{\alpha_m}}  \right) ,
\label{HD_rho_1}
\end{align}
where the subscripts $S$ indicating the signal mode have all been replaced with $R$ indicating the received mode.

Equation \ref{HD_rho_1} allows for direct comparison with the corresponding density operator for the previous spoofing strategy (cf. Eq. \ref{DD_rho_1}). However, since the idler and received modes in this case are in Gaussian states, it is convenient to represent the them with covariance matrices. The states given in Eqs. \ref{DD_rho_0} and \ref{HD_rho_1} have the corresponding covariance matrices
\begin{align}
    V_{H_0}&=\left ( \begin{array}{cccc}
     \omega & 0& c & 0\\
    0 & \omega&0 &-c\\
     c&0 & \nu &0 \\
     0&-c & 0&  \nu
\end{array}\right),
\label{covariance_null_hyp}
\end{align}
where $\nu = N+1/2$, $\omega = \tau N + N_o +1/2$, $c=\sqrt{\tau N(N+1)}$, and
\begin{align}
    V_{H_1}&=V_{H_0} + \left ( \begin{array}{cccc}
     \sqrt{\tau} & 0& 0 & 0\\
    0 & \sqrt{\tau}&0 &0\\
     0&0 & 0 &0 \\
     0&0 & 0& 0
\end{array}\right),
\label{covariance_alt_hyp}
\end{align}
respectively. The second term on the right in Eq. \ref{covariance_alt_hyp} represents the one photon of quantum noise introduced by the spoofer's measure-and-prepare strategy, re-scaled by the loss in propagation to the receiver. This quantum noise can be viewed as breaking the entanglement between the idler and received modes. 

The fidelity of these two zero-mean, two-mode Gaussian states, expressed in terms of the parameters of the covariance matrices, is \cite{marian2012uhlmann, banchi2015quantum}
\begin{align}
    F(\hat{\rho}_{H_0}, \hat{\rho}_{H_1})=\frac{1}{\sqrt{\left(\sqrt{\Gamma}+\sqrt{\Lambda}\right)-\sqrt{\left(\sqrt{\Gamma}+\sqrt{\Lambda}\right)^2-\Delta}}},
\end{align}
where 
\begin{align}
    \Delta &=\left[2\nu(2\omega+\sqrt{\tau})-4c^2\right]^2, \\
    \Gamma =& 16\left([\omega\nu-c^2][(\omega+\sqrt{\tau})\nu-c^2]+ \right .\nonumber \\
    &\left .\frac{1}{4}(\omega(\omega+\sqrt{\tau})+\nu^2-2c^2)+\frac{1}{16} \right)^2,
\end{align}
and
\begin{align}
    \Lambda &= 16\left\{ (\omega\nu-c^2)^2-\frac{1}{4}[\omega^2 +\nu^2 -2c^2]+\frac{1}{16}\right\} \\
    &\times \left\{((\omega+\sqrt{\tau})\nu-c^2)^2-\frac{1}{4}[(\omega+\sqrt{\tau})^2 +\nu^2 -2c^2]+\frac{1}{16}\right\} \nonumber
\end{align}

An example of the fidelity as a function of signal-to-noise ratio is shown in Fig. \ref{fig:fidelity_in_noise} (dashed red line), where $N = 0.01$, $M=10^4$, and $\tau = 10^{-6}$. The fidelity with heterodyne detection and coherent state generation is lower than the fidelity for direct detection and number state generation over the full range. We conclude that the advantage of the latter strategy over the former in the noise-free, loss-free model above survives the introduction of these effects.

\section{Conclusion}
Our results indicate entanglement in the QI signal is a resource for detecting the presence of a spoofer who employs a measure-and-prepare strategy. The utility of this resource should be weighed in future assessments of practical applications for quantum radar \cite{torrome2023advances,luong2023quantum,bischeltsrieder2024engineering}. The optimal strategy for such a spoofer to avoid detection remains an open question. We examined two specific spoofing strategies and found both are, in principle, susceptible to detection. The ability to detect spoofing increases with signal brightness and bandwidth. Notably, this ability continues to increase with brightness beyond the weak signal regime where QI has a quantum advantage. The most effective of the two spoofing methods analyzed here is mode-wise direct detection with multi-mode number state generation. Since no measure-and-prepare strategy can produce a received mode that is entangled with a remote idler mode, we expect that any classical spoofing strategy has some degree of vulnerability. It is interesting to contemplate non-classical strategies the spoofer might employ.

\appendix*
\section{Effect of Noise-loss Channel on One Mode of a Two-mode State}
\label{Noise-Loss Channel}
Suppose one mode of a two-mode state is input to a noise-loss channel. Let the initial density operator be written in number state representation as
\begin{align}
    \hat{\rho}= \sum_{i=0}^\infty\sum_{j=0}^\infty\sum_{k=0}^\infty\sum_{l=0}^\infty \rho_{ij,kl} \ket{i}_I \ket{j}_S\prescript{}{I}{\bra{k}}\prescript{}{S}{\bra{l}},
\end{align}
where $I$ and $S$ indicate idler and signal, respectively. Let the signal mode pass through the channel $\mathcal{L}_{\tau, N_i}^{S}$. The action of the channel is equivalent to the process of combining the signal mode on a beam splitter of transmissivity $\tau$ with a thermal state of mean photon number $N_i$ and tracing over one output mode. The resulting two-mode state is
\begin{align}
     &\mathcal{L}_{\tau, N_i}^S\left(\hat{\rho}\right)=\sum_{j=0}^\infty\sum_{k=0}^\infty\sum_{l=0}^\infty\sum_{m=0}^\infty\rho_{lj,mk}\sum_{n=0}^\infty \frac{N_i^n}{(1+N_i)^{n+1}}\frac{1}{n!\sqrt{j!k!}}\ket{l}_I\prescript{}{I}{\bra{m}}\;\otimes\nonumber \\
      &  \sum_{r=0}^j \sum_{s=0}^n\binom{j}{r}\binom{n}{s} (\sqrt{\tau})^{n-s+r}     (\sqrt{1-\tau})^{j-r+s}(-1)^{n-s}  \sqrt{(j-r+n-s)!} \sqrt{(r+s)!}   \times \nonumber \\
     & \sum_{t=0}^k \sum_{u=0}^n\binom{k}{t} \binom{n}{u}(\sqrt{\tau})^{n-u+t}   (\sqrt{1-\tau})^{k-t+u} (- 1)^{n-u} \sqrt{(k-t+n-u)!} \sqrt{(t+u)!} \times \nonumber \\
     &\delta_{j-r+n-s,k-t+n-u}\ket{r+s}_1 \prescript{}{1}{\bra{t+u}}
\end{align}

\input{main.bbl}
\end{document}

%% file: main.bbl
%